\documentclass[letterpaper,10pt, conference]{ieeeconf}      

\IEEEoverridecommandlockouts                              
\usepackage{amsmath,amssymb,amsfonts,theorem}
\usepackage{graphicx,color,psfrag}
\usepackage{subfigure}
\graphicspath{{./Figures/}}

\def\qedp{\hspace*{\fill}~{\tiny $\blacksquare$}}

{ \theorembodyfont{\normalfont} 

\newtheorem{remark}{Remark}
}
\newtheorem{assumption}{Assumption}
\newtheorem{definition}{Definition}
\newtheorem{theorem}{Theorem}
\newtheorem{lemma}{Lemma}

\def\QEDclosed{\mbox{\rule[0pt]{1.3ex}{1.3ex}}} 
\def\QEDopen{{\setlength{\fboxsep}{0pt}\setlength{\fboxrule}{0.2pt}\fbox{\rule[0pt]{0pt}{1.3ex}\rule[0pt]{1.3ex}{0pt}}}}
\def\QED{\QEDopen}

\providecommand{\com[2]}{\begin{tt}[#1: #2]\end{tt}}

\def\be{\begin{equation}}
\def\ee{\end{equation}}
\def\ba{\begin{array}}
\def\ea{\end{array}}
\def\eqa{\begin{eqnarray}}
\def\eqe{\end{eqnarray}}

\title{On Resilient Control of Nonlinear Systems under Denial-of-Service}
\author{Claudio De Persis  \and  Pietro Tesi
\thanks{C. De Persis and Pietro Tesi are with Faculty of Mathematics and Natural Sciences, University of Groningen, the Netherlands, {\tt \{c.de.persis,p.tesi\}@rug.nl}. This work 
is partially supported by the Dutch Organization for Scientific
Research (NWO) under the auspices of the project QUICK (QUantized
Information Control for formation Keeping).}
}
\begin{document}

\maketitle


%
\begin{abstract}
We analyze and design a control strategy for nonlinear systems under Denial-of-Service attacks. Based on an ISS-Lyapunov function analysis, we provide a characterization of the maximal percentage of time during which feedback information  can be lost without resulting in instability of the system. Motivated by the presence of a digital channel we consider event-based controllers for which a minimal inter-sampling time is explicitly characterized.  
\end{abstract}

\section{Introduction} \label{sec:introduction}

Motivated by interest in the analysis and control of critical infrastructures such as power networks, supply chains and transportation systems, recent years have witnessed increasing research interests in large-scale engineered systems. To achieve the prescribed control goal, these systems require exchange of information that often occurs in digital form. In turn this has triggered interest in control over communication channels. One of the topics that has stimulated broad interest is the so-called event-based control (\cite{Tabuada07}) in which sampling times are designed in real-time with the ultimate goal of saving communication resources while still guaranteeing the control goal. Event-based control has found fertile ground also in the area of cooperative control; \emph{e.g.}, see \cite{GS-DVD-KHJ:13, CDP:PF:TAC13}. \vspace{0.1cm}

A natural research question raises when dealing with control over a communication channel: 
whether or not stability properties and performance are preserved in the presence of loss of feedback information.
This loss of information could be due not only to malfunctioning but also to malicious actions by an adversarial entity \cite{sastry,sinopoli}. In the latter case, the assumption on the kind of information loss should be kept to a minimum since intelligent adversaries might not follow e.g.~any statistical pattern. This aspect is in contrast with other work  where the loss of information is mainly due to the unreliability of the communication channel \cite{Schenato}. \vspace{0.1cm}
  
Several contributions to the topic of stability/stabilization in the presence of adversarial entities 
have been reported in the last few years, with main emphasis on the so-called 
\emph{Denial-of-Service} (DoS), a class of attack strategies primarily intended to affect the timeliness of information exchange \cite{Lowe}.
In \cite{sastry}, the authors address 
the problem of security constrained optimal control for discrete-time linear 
systems in which packets may be jammed by a malicious adversary, and 
the goal is to find optimal control and attack strategies 
assuming a maximum number of jamming actions over a prescribed (finite) control horizon.
A very similar scenario is considered in \cite{basar}, where 
the problem of stabilizing a discrete-time linear system under DoS 
is casted as a dynamic zero-sum game.
An interesting alternative scenario is addressed in \cite{martinez},
where the authors consider the problem of stability under \emph{periodic} DoS
for linear sampled-data systems under state-feedback. The idea there is to identify 
the jamming signal so as to restrict the information exchange to the time 
intervals where no DoS occurs. This approach has been then extended in \cite{HSF-SM:13-ijrnc} 
by considering energy-constrained, but otherwise \emph{unknown} DoS attacks. \vspace{0.1cm}

In \cite{CDP:PT:IFAC14,CDP:PT:ARXIV}, we addressed afresh 
the problem of stability under energy-constrained, but {unknown}, DoS attacks
within the framework of linear sampled-data systems under state-feedback. 
%
The analysis differs from the one in \cite{HSF-SM:13-ijrnc,martinez}
since the goal is not to identify the jamming signal; rather, the goal is
to determine if stabilization is possible assuming only a bound on the fraction
of the time the jammer is active. The considered approach, inspired by \cite{Tabuada07}, consists in  
a suitable logic that determines in real-time the frequency 
of controller updates (the sampling times)
depending on the DoS occurrence.
In particular, it enjoys the following 
features: \vspace{0.1cm}

i) It ensures \emph{global exponential} stability of the closed-loop system
whenever the intervals over which 
communication is possible are {predominant} with respect to
the intervals over which communication is denied; \vspace{0.1cm}

ii) It allows for the state-feedback matrix to be designed in accordance with any control design method,
robustness against DoS being achieved thanks 
to the sampling logic; \vspace{0.1cm}

iii) It is \emph{resilient} since the sampling rate varies
depending on the DoS occurrence; \vspace{0.1cm}

iv) It allows for an explicit characterization of 
convergence rate, minimal inter-sampling time, and
ratios between the ``active'' and ``sleeping'' periods
of DoS which do not destroy closed-loop stability; \vspace{0.1cm}

v) It is flexible enough so as to allow the designer to choose 
from several implementation options that can be used to trade-off performance vs. communication resources. \vspace{0.1cm}

The objective of this paper is to initiate the investigation of similar ideas for nonlinear systems. 
Although we follow the line of arguments of \cite{CDP:PT:IFAC14,CDP:PT:ARXIV}, 
a few of the steps we take are very peculiar to nonlinear systems, making the 
extension  far from straightforward and deserving attention on its own right. 
It is shown that under certain additional conditions, 
which are basically needed to avoid finite-escape times phenomena 
during DoS, \emph{semi-global asymptotic stability} can be still 
ensured. The analysis combines arguments from event-based control and
ISS control Lyapunov functions. \vspace{0.1cm}

The remainder of this paper is organized as follows. In Section \ref{sec.problem} 
we introduce the framework of interest and provide an overview of the problem.
In Section \ref{sec.prep}, we describe 
the considered class of DoS attacks and provide some preliminary stability results.
The main result  with a characterization of the class 
of DoS signals under which stability is preserved is given in Section \ref{sec:switch}. 
In Section \ref{sec.finite.rate}, we provide a characterization 
of the achieved minimal inter-sampling rate.
Section \ref{sec:conclusions} provides concluding remarks
and outlines future research directions.

\section{Framework and problem overview}\label{sec.problem}

In this paper we consider nonlinear systems of the form 
\be\label{system} 
\dot x= f(x,u)
\ee
for which we assume the existence of a smooth state feedback $u=k(x)$ which renders the closed-loop system
\be
\dot x= f(x,k(x+e)) \nonumber
\ee
ISS with respect to measurement errors $e$ in the sense that there exist a smooth function $V$ and class ${\mathcal K}_\infty$ functions $\alpha_1, \alpha_2, \gamma_2$ such that 
\be\label{iss}
\ba{l}
\alpha_1(\| x \|) \le V(x)\le \alpha_2(\| x\|)\\\\
\nabla V(x) f(x, k(x+e))\le -\lambda V(x) +\gamma_2 (\| e \|),
\ea
\ee 
with $\lambda>0$. \vspace{0.1cm}

The control action is implemented 
via a \emph{sample-and-hold} device. In a {\em nominal} situation, given a sequence of times 
$\{t_k\}$, $k \in \mathbb N$, where $t_0:=0$ by convention,  the control action is such that 
\be\label{ideal_control}
u_{{\rm nom}}(t)= k(x(t_k)), \quad \mbox{for all } t\in [t_k, t_{k+1}). 
\ee
The mechanism that generates this sequence of times will be specified in the sequel. 
By nominal situation is meant that at each time $t_k$ at which the actuator needs to update the control value, it correctly receives the sampled value $k(x(t_k))$. \vspace{0.1cm}

The focus of this paper is on a scenario that is different from the nominal one, namely one in which there might be times in the sequence $\{t_k\}$ at which the control value cannot be updated because no information regarding $k(x(t_k))$ is received by the actuator. This loss of information can be caused by several factors, such as a defective communication channel or as a consequence of the action of an adversarial entity. To fix the ideas, we focus in the sequel on the latter scenario, and refer to this interruption of information transmission from the sensor to the actuator  as {\em Denial of Service} or DoS. \vspace{0.1cm}

Let 
$\{h_n\}$, $n \in \mathbb N$, $h_0 \geq 0$, represent a  sequence of ``positive edge-triggering'' times which define the time intervals at which a DoS attack is occurring. Namely, at $h_n$ the $n$-th DoS attack becomes active (no communication is possible from the sensors to the actuators), while at $h_n+\tau_n$, with $\tau_n >0$ the duration of the  $n$-th DoS attack during which information transmission is not possible, the DoS attack ends (communication is possible from the sensors to the actuators). In formula,  
\begin{equation}  \label{DoS_intervals}
H_n = [h_n,h_n+\tau_n[  
\end{equation}
represents the $n$-th  DoS time-interval. \vspace{0.1cm}


\begin{figure}[tb]
\begin{center}
\psfrag{x}{{\tiny $x$}}
\psfrag{u}{{\tiny $u$}}
\includegraphics[width=0.45 \textwidth]{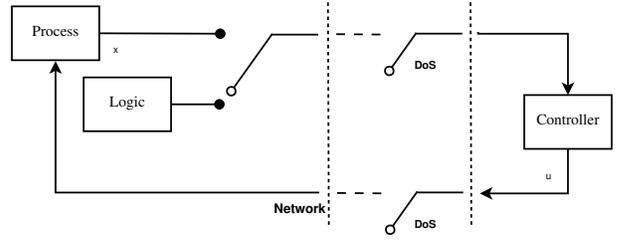} 
\caption{Block diagram of the closed-loop system 
under DoS on the communication channels.} \label{fig:BasicScheme}
\end{center}
\end{figure}

We then assume that, in the presence of DoS,  
the actuator generates an input that is 
based on the \emph{most recently received} control signal. 
Specifically, denote the set of time-instants 
where communication is possible by
\begin{eqnarray}  \label{Theta}
\Theta(t) := \left[0,t \right]  \setminus \bigcup_{n \in \mathbb N} H_n 
\end{eqnarray}
where $\setminus$ means relative complement.
Accordingly, the control input applied to the process at each time instant
can be expressed as
\begin{eqnarray}  \label{actual_control}
u(t) = k(x(t_{k(t)}))
\end{eqnarray}
where
\begin{eqnarray}  \label{last_succesful_control_update}
k(t) := \left\{ 
\begin{array}{ll} 
-1, & \textrm{if} \,\, \Theta(t) = \emptyset  \\  \\
\begin{array}{l}
\sup \, \left\{ \, k \in \mathbb N\, | \,\, t_k \in \Theta(t)  \,   \right\}, 
\end{array}
& \textrm{otherwise}
\end{array}
\right.
\end{eqnarray}
denote the last (up to the current time) successful control update.
Notice that $h_0=0$ implies $k(0)=-1$, which raises 
the question of assigning a value to the control input 
when communication is not possible 
at the process start-up. In this respect, we assume that 
when $h_0=0$ then $u(0) = 0$,  and we let $x(t_{-1}) := 0$
for notational consistency. \vspace{0.1cm}

%
%


The problem of interest is to find suitable control update rules,
\emph{i.e.} logics for generating $\{t_k\}_{k \in \mathbb N}$
that are \emph{resilient} 
against DoS,
while ensuring the existence of a minimal 
control update inter-execution time.  
The requirement of resilience calls for logics capable not only 
to tolerate but also to counteract the occurrence of DoS.
On the other hand, requiring the existence of a minimal 
control update inter-execution time is fundamental 
in order for the control architecture to be implementable 
on digital platforms. \vspace{0.1cm}

The following definitions are key for the developments 
of the paper.



\begin{definition} \label{def:GES}
System
$\Sigma$,  composed of 
(\ref{system}) in closed-loop with 
(\ref{actual_control}) is said to have a  \emph{semi-globally asymptotically stable}  origin 
if there exists a $\mathcal{K}\mathcal{L}$ function $\beta$ such that, for all $R>0$, the solution to $\Sigma$ satisfies 
\begin{equation} \label{GES}
\| x(t) \| \leq \beta(\| x(0) \|, t)
\end{equation}
for all $t \in \mathbb R_{\geq 0}$ and for all $x(0) \in \mathbb R^{n_x}$ such that $\|x(0)\|\le R$,
where $\|\cdot\|$ stands for Euclidean norm. 
\hfill $\Box$
\end{definition}


\begin{definition} \label{def:FSR}
A control update sequence $\{t_k\}$ 
is said to occur at a \emph{finite sampling rate} if
there exists an $\varepsilon \in \mathbb R_{>0}$
such that 
\begin{eqnarray} \label{minimal_IET}
 \Delta_k \, :=\,  t_{k+1} -t_k \, \geq \, \varepsilon 
\end{eqnarray}
for all $k \in \mathbb N$.
\hfill $\Box$
\end{definition}

\begin{remark}
It is worth pointing out that the synthesis of control laws achieving ISS is in general a nontrivial task. 
Nonetheless, the literature on the topic is quite vast. 
In particular, contributions centered around the concept of ISS-control Lyapunov functions
can be found in \cite{Freeman96,Tsinias97,Liberzon02}; see also \cite{Vu07} for relevant design examples. \hfill $\Box$
\end{remark}

\section{A preparatory analysis}\label{sec.prep} 

In this section, we make some considerations 
regarding the considered class of DoS attacks and provide some preliminary stability results.

\subsection{DoS attacks}

In general the uncontrolled system (\ref{system}) (i.e.~with $u=0$) might have an unstable dynamics 
and also exhibit finite-escape times. This suggests two facts: (i) the duration of a DOS attack cannot be 
arbitrarily large; and (ii) conditions on the systems and of the DoS attack must make sure that 
information is delivered before the occurrence of a possible finite-escape time.   \vspace{0.1cm}

Given a sequence $\{h_n\}$, let
\begin{eqnarray}  \label{DoS_intervals_union}
\Xi(t) \, := \,  \bigcup_{n \in \mathbb N} H_n  \, \bigcap  \, [0,t]
\end{eqnarray}
denote the total interval of DoS up to the current time,
where, given an interval $I$, let $|I|$ denote its length. 
The following assumptions are then considered.


\begin{assumption} \label{ass:DoS_slow}
The DoS sequence $\{h_n\}$, $n \in \mathbb N$,  is such that 
$\lim_{n \rightarrow \infty } h_n =\infty$. Moreover,
there exist constants $\kappa \in \mathbb R_{\geq0}$ and $\tau \in \mathbb R_{>0}$ such that
\begin{equation}  \label{DoS_slow} 
|\Xi(t)| \, \leq \, \kappa + \frac{t}{\tau}
\end{equation}
for all $t \in \mathbb R_{\geq 0}$. 
\hfill $\Box$
\end{assumption}

\begin{assumption}\label{assumpt2}
There exists a positive value $\mu$ such that 
\be\label{bound.assumpt2}
\gamma_2(4r)\le \mu \alpha_1(r)
\ee 
for all $r\in \mathbb{R}_{\ge 0}$. \hfill $\Box$
\end{assumption}

In accordance with (i), limiting the duration of DoS is 
necessary for closed-loop stability to be conceptually 
achievable. Assumption \ref{ass:DoS_slow} captures well this prescription
as it requires the existence of a bound on the \emph{fraction
of time} in which DoS is active. On the other hand, 
no conditions are imposed on the DoS ``structure'': first,
DoS is allowed to occur aperiodically; second, the duration 
of any two different DoS intervals need not be equal to one another. \vspace{0.1cm}

On the other hand, the role of Assumption \ref{assumpt2} is that of
limiting the rate of divergence of the process states during the 
time intervals over which the control action cannot be updated.
In this connection, as will become apparent in the sequel, Assumption \ref{assumpt2}  
ensures that over such intervals the rate of divergence of the process states
is at most exponential, which guarantees the absence 
of finite-escape times phenomena. Notice that when the 
process dynamics are linear, Assumption \ref{assumpt2} 
is always satisfied.


%

\subsection{Ideal sampling logic and preliminary analysis}

Given Assumptions  \ref{ass:DoS_slow} and \ref{assumpt2}, one expects that 
stability is not destroyed if the intervals over which 
communication is possible are \emph{predominant} with respect to
the intervals over which communication is denied. 
However, proving this fact is far from straightforward within the classical framework 
of nonlinear sampled-data systems (\cite{Nesic,Nesic2,Nesic3,Bian}).
Even more importantly, \emph{periodic} sampling strategies hamper 
the possibility of increasing the closed-loop robustness  by adapting the sampling rate to the DoS occurrence.
As shown next, the use of \emph{aperiodic} sampling strategies as
introduced by \cite{Tabuada07}
provides a convenient framework to work with, in terms of both 
ease of analysis and effectiveness. \vspace{0.1cm}

Consider the system $\Sigma$ composed of 
(\ref{system}) in closed-loop with 
(\ref{actual_control}), namely
\[
\dot x(t)= f(x(t), k(x(t_{k(t)})), \quad t\in [t_{k(t)}, t_{k(t)+1}). 
\]
Let
\begin{equation} \label{error}
e(t) := x(t_{k(t)}) - x(t) 
\end{equation}
be the measurement error induced by the sampling, with $e(t_{k(t)})=0$
by definition. System $\Sigma$  can be therefore written as
\[
\dot x(t)= f(x(t), k(x(t)+e(t))), \quad t\in [t_{k(t)}, t_{k(t)+1}). 
\]
Consider next the following \emph{ideal} sampling logic.
Given a sampling instant $t_k$: \vspace{0.1cm}

i) if $t_k$ does not belong to $\Xi(t)$, then the next sampling 
instant $t_{k+1}$ is defined as the {infimal} time larger than $t_{k}$ such that the condition 
\be\label{cur}
\gamma_2(4 \|e(t)\|) \le \lambda(1-c) V(x(t)), 
\ee
with $c\in (0,1)$, is violated; \vspace{0.1cm}

ii) if instead $t_k$ belongs to some 
DoS interval $H_n$,
then the next sampling instant $t_{k+1}$ is defined as  
$h_n+\tau_n$. \vspace{0.1cm}


This logic ensures that ({\ref{cur}}) holds true for all $t \in \Theta(t)$. 
Unfortunately, it is \emph{ideal} in the sense that it cannot be implemented
over digital platforms. In fact, because of the possible
aperiodic nature of DoS, it is conceptually impossible 
to foresee when Dos will cease, and, hence, it is impossible 
to implement point ii). Nonetheless, at this stage of analysis, it is convenient 
to focus on this logic. Variants that are implementable
over digital platforms will be discussed in Sections \ref{sec:switch} and \ref{sec.finite.rate}. \vspace{0.1cm}





 
We are now in the position to provide a preliminary closed-loop stability analysis. 
The underlying idea can be described as follows: 
we decompose the time axis into the sets $\Theta(t)$ and $\Xi(t)$.
Under the considered logic,  (\ref{cur}) holds by construction
over $\Theta(t)$. In turns, this ensures that (\ref{iss}) 
satisfies a dissipation-like inequality.
On the other hand, over $\Xi(t)$, (\ref{cur}) need not hold
and the closed-loop system might exhibit divergence trends.
Overall, the closed-loop dynamics can be therefore viewed as those
of a hybrid system, where one switches between stable and unstable modes. The idea is then to determine 
conditions under which the stable behavior is predominant with respect to the unstable one. \vspace{0.1cm}



We make these considerations precise. 
First notice that, as an immediate consequence of (\ref{cur}), for all $t  \in \Theta(t)$,
we have
\[
\dot V(x(t))\le - c\lambda V(x)=: -\omega_1 V(x(t)).
\] 
On the other hand, for all $t\in H_n=[h_n, h_n+\tau_n]$ for some $n\in \mathbb{N}$, we have that 
\[
\dot V(x(t))  \le -\lambda V(x(t)) +\gamma_2 (\| e(t)\|).
\] 
We would like to bound the growth of $e(t)$ as a function of $V(x(t))$. Recall that 
\[
\| e(t)\|\le \| x(t_{k(h_n)})- x(t)\|\le \| x(t_{k(h_n)})\| + \|x(t)\|. 
\]
To bound $\| x(t_{k(h_n)})\|$, we need to introduce an appropriate lemma. 

\begin{lemma}\label{lem:tech:nonl}
Under Assumption \ref{assumpt2}, the state $x(t)$ of system (\ref{system}) under the control update rule (\ref{cur}) and sample-and-hold feedback control (\ref{actual_control}) satisfies  
\be\label{bound:nonl}
\ba{rcl}
\| x(t_{k(h_n)})\| &\le& \displaystyle\frac{1}{4} \gamma_2^{-1}(\lambda(1-c)V(x(h_n)))+\\[4mm]
&& \displaystyle\frac{1}{4}   \gamma_2^{-1}(\mu V(x(h_n)))
\ea\ee
over the domain of existence of the solution $x$. 
\end{lemma}
The proof of the result is given in the Appendix. \vspace{0.1cm}

Notice that 
\[\ba{c}
\gamma_2(\| e(t)\|) \le 
\gamma_2(2\| x(t_{k(h_n)})\|)+\gamma_2(2\| x(t)\|)
\ea\]
where we have exploited the inequality $\gamma_2(a+b)\le \gamma_2(2a)+\gamma_2(2b)$. From (\ref{bound:nonl}), we obtain
\[\ba{c}
\gamma_2(2\| x(t_{k(h_n)})\|) \le
\gamma_2\left(\frac{1}{2} \gamma_2^{-1}(\lambda(1-c)V(x(h_n)))+
\right.\\ \\
\left.\frac{1}{2}   \gamma_2^{-1}(\mu V(x(h_n)))\right)\le 
(\lambda(1-c)+\mu)V(x(h_n))
\ea\]
On the other hand 
\[
\gamma_2(2\| x(t)\|) \le \gamma_2(2\alpha_1^{-1}(V(x(t))))\le \mu V(x(t))
\]
where the last inequality descends from (\ref{bound.assumpt2}).\\
Hence, for all $t\in H_n$, we have 
\[\ba{c}
\dot V(x(t))  \le -\lambda V(x(t)) +\gamma_2 (\| e(t)\|)\\ \\
\le (\mu-\lambda) V(x(t)) +(\lambda(1-c)+\mu)V(x(h_n)).
\ea\] 
For those $t\in H_n$  such that $V(x(h_n))\le V(x(t))$,
\be\label{lyap_bound_4}\ba{c} 
\dot V(x(t)) 
\le (2 \mu-c\lambda) V(x(t))<\omega_2 V(x(t)),
\ea\ee 
with $\omega_2:=\lambda(1-c)+2\mu$,
while for those $t\in H_n$  such that $V(x(h_n))> V(x(t))$,  
\be\label{lyap_bound_5}\ba{c}
\dot V(x(t))  
\le (\lambda(1-c)+2\mu)V(x(h_n))= \omega_2 V(x(h_n)).
\ea\ee 
Inequalities (\ref{lyap_bound_4}) and (\ref{lyap_bound_5}) can be combined to prove the following: 

\begin{theorem} \label{thm:gb2}
Consider the control 
system $\Sigma$ composed of 
(\ref{system}) in closed-loop with 
(\ref{actual_control}), under DoS attacks satisfying Assumption  \ref{ass:DoS_slow},
and with control update rule defined in (\ref{cur}). Let  Assumption \ref{assumpt2} hold. \vspace{0.1cm}

If the parameter $\tau$ in (\ref{DoS_slow}) satisfies
\begin{equation} \label{tau_gb2}
\tau \, > \,  \frac{\omega_1}{\omega_1+\omega_2}=\frac{c\lambda}{\lambda+2\mu},  
\end{equation}
where 
\[
\omega_1=c\lambda,\quad \omega_2=\lambda(1-c)+2\mu,
\]
then any solution to $\Sigma$ satisfies the inequality 
\[
\| x(t)\|\le \alpha_1^{-1}\left(
e^{\kappa(\omega_1+\omega_2)} \, e^{-[\omega_1 -  (\omega_1+\omega_2)/\tau]\,t} \alpha_2(\| x(0)\|)
\right)
\]
for all $t\ge 0$. 
%
\end{theorem}

We refer the reader to the Appendix for a proof of the theorem.

\begin{remark}
The result proves {\em global} asymptotic stability. In fact, the need to restrict the set of initial conditions within a ball of arbitrary large radius, thus obtaining a {\em semi-global} asymptotic stabilizability result, becomes apparent when we prove in Section \ref{sec.finite.rate} that the sequence of control update times generated by (\ref{cur}) occurs at a finite rate. 
\end{remark}

\section{DoS-induced actuation delay}\label{sec:switch}

The analysis of the previous section  rests upon the fulfillment of 
condition (\ref{cur}). In practice the sequence of control update times  occurs at a finite sampling rate. This has a consequence on the actual duration of the DoS. As a matter of fact, when a sensor attempts to transmit and no acknowledgement is received due to the attack, it will repeat the transmission attempts until the transmission is successful. Due to the finite transmission rate, even when transmission becomes possible, there will be a delay from the time the DoS attack is over and the time the transmission can successfully occur. This delay causes a prolongation of the DoS interval that affects the stability result of the previous section. In this section, after recalling some notation from \cite{CDP:PT:IFAC14,CDP:PT:ARXIV}, we provide a stabilization result that takes into account such a prolongation of the DoS attack.  In other words, we remove the second simplifying assumption stated in ii) after (\ref{cur}). \vspace{0.1cm}

Consider  a control update sequence $\{t_k\}$ along with a DoS sequence $\{h_n\}$,
and let
\begin{eqnarray} \label{}
\mathbb S_n \, := \,  \left\{ k \in \mathbb N \, |\,\, t_k \in H_n \right\}
\end{eqnarray} 
denote the set of integers associated with an attempt 
to update the control action during $H_n$. Accordingly,
by defining
\begin{eqnarray} \label{}
\Delta_{\mathbb S_n} \, := \, \sup_{k \in \mathbb S_n}  \Delta_k
\end{eqnarray} 
then
\begin{eqnarray} \label{DoS_interval_plus_FSR}
\bar H_n := [h_n,\, h_n + \tau_n + \Delta_{\mathbb S_n}[
\end{eqnarray}  
will provide an upper bound on the $n$-th time interval over which 
the control action is not updated, while
\begin{eqnarray}  \label{DoS_intervals_union_digital}
\bar \Xi(t)  \, := \,  \bigcup_{n \in \mathbb N} \bar H_n  \, \bigcap  \, [0,t]
\end{eqnarray}
will provide an upper bound on the total interval up to the current time over which 
the control action is not updated. Equation (\ref{DoS_interval_plus_FSR})
essentially models the additional delay in the control update that 
may arise under finite sampling rate. \vspace{0.1cm}

We are now ready to state a version of  Theorem \ref{thm:gb2} in which the DoS-attack-induced delay is taken into account.

\begin{theorem} \label{thm:gb_2}
Consider the control 
system $\Sigma$ composed of 
(\ref{system}) in closed-loop with 
(\ref{actual_control}), under DoS attacks satisfying Assumption  \ref{ass:DoS_slow},
and with control update rule defined in (\ref{cur}). 
Let  Assumption \ref{assumpt2} hold. 
\\
If the parameter $\tau$ in (\ref{DoS_slow}) satisfies
\begin{equation} \label{tau_gb2_mod}
\tau \, > \,  \frac{c\lambda}{\lambda+2\mu}\left( 1+ \frac{\Delta_*}{\tau_*} \, \right),  
\end{equation}
where
\begin{equation} \label{Delta_*}
\Delta_* \, := \, \sup_{n \in \mathbb N}  \Delta_{\mathbb S_n}
\end{equation}
and 
\begin{equation} \label{tau_*}
\tau_* \, := \, \inf_{n \in \mathbb N}  \tau_{n}>0,
\end{equation}
then  the inequality 
\be\label{nonl.bound.x}
\| x(t)\|\le \alpha_1^{-1}\left(\gamma {\rm e}^{-\beta t} \alpha_2(\| x(0)\|)
\right)
\ee
holds for all $t\ge 0$, where
\[
\gamma =e^{\kappa (\lambda+2\mu)\left( 1+ \frac{\Delta_*}{\tau_*} \, \right)},\quad  \beta= \left(c\lambda- \frac{\lambda+2\mu}{\tau}\left(1+ \frac{\Delta_*}{\tau_*}\right)\right)
\]
and the parameters $\kappa, \lambda, \mu$,  are as in (\ref{DoS_slow}), (\ref{iss}) and (\ref{bound.assumpt2}), respectively. 
\end{theorem}

\emph{Proof}. For any $t\in \mathbb{R}_{\ge 0}$, the Lyapunov function evolves as $\dot V(x(s))\le -\omega_1 V(x(s))$ if $s\not \in \overline{\Xi}(t)$   and as 
$\dot V(x(s))\le \omega_2 V(x(s))$ if $s\in \overline{\Xi}(t)$. Hence
\[\ba{rcl}
V(x(t)) 
&\le & {\rm e}^{-\omega_1 t + (\omega_1+\omega_2)\left| \overline{\Xi}(t)\right|}  V(x(t_0)).
\ea\]
In \cite{CDP:PT:ARXIV}, Theorem 2, the following estimate of $\left| \overline{\Xi}(t)\right|$ is provided:
\begin{eqnarray} \label{}
|\bar \Xi(t)| \,&\leq& \,  \left( \kappa+ \frac{t}{\tau} \, \right)  \left( 1+ \frac{\Delta_*}{\tau_*} \, \right) 
\end{eqnarray}
Hence,  
\footnotesize
\[\ba{rcl}
V(x(t)) \le  {\rm e}^{-\left(\omega_1- \frac{\omega_1+\omega_2}{\tau}\left(1+ \frac{\Delta_*}{\tau_*}\right)\right) t + \kappa (\omega_1+\omega_2)\left( 1+ \frac{\Delta_*}{\tau_*} \, \right)}  V(x(t_0)).
\ea\]
\normalsize
The thesis now follows immediately under the stated assumptions and bearing in mind that $\omega_1= c\lambda, \omega_2=\lambda(1-c) +2\mu$.

 \hfill $\Box$


\begin{remark}  \label{rem:thm_2}
Theorem \ref{thm:gb_2} differs from Theorem \ref{thm:gb2} not only 
because of $\Delta_*$ but also due to the presence of $\tau_*$.
This has a very intuitive explanation. In fact, in the ideal case considered in Theorem \ref{thm:gb2}, 
$\Delta_*=0$ since a control update
can always occur as soon as DoS is over. 
Under finite sampling rate, each DoS interval will instead 
possibly introduce an additional delay in the control update.
This also points out that, given two DoS sequences 
of equal total length, the one composed of more intervals 
having smaller duration will be more critical for stability,
since it will potentially deny more communications attempts.
\hfill $\Box$
\end{remark}

\section{Finite sampling rate}\label{sec.finite.rate}

Until now in the investigation we have been implicitly assuming that the sampling  sequence generated by the rule (\ref{cur})  occur at a finite sampling rate. 
In this section we investigate conditions under which such assumption is actually met in practice. To this end, we restrict the set of initial conditions for the process. In fact, we say that the solutions to the control 
system $\Sigma$ composed of 
(\ref{system}) in closed-loop with 
(\ref{actual_control}), under DoS attacks satisfying Assumption  \ref{ass:DoS_slow},
and with control update rule defined in (\ref{cur}),  have (\cite{RP-PT-DN-AA-CDC11})   a {\em semi-global uniform finite sampling rate} if for every $R>0$ and for every DoS sequence  satisfying Assumption \ref{ass:DoS_slow}, there exists $\varepsilon_R>0$ such that any solution to $\Sigma$ starting in the ball $B_R(0)$ of radius $R$ and center the origin has an 
  associated control update   sequence $\{t_k\}$ that for all $k \in \mathbb N$ satisfies (\ref{minimal_IET}) with $\varepsilon$ replaced by $\varepsilon_R>0$. \vspace{0.1cm}  

%

Let $R$ be an arbitrary positive number such that $\|x(0)\||\le R$. Motivated by the bound in (\ref{nonl.bound.x}), we introduce the balls 
\[
\mathcal{X}= B_{\alpha_1^{-1} (\gamma \alpha_2(R))}(0)\subset \mathbb{R}^n,\; 
\mathcal{E}= B_{2\alpha_1^{-1} (\gamma \alpha_2(R))}(0)\subset \mathbb{R}^n\; .
\]
Notice that these balls are known {\it a priori} since their radius depend on quantities that are fixed in advance. \\
Assume that the vector field $f(x, k(x+e))$ is locally Lipschitz with respect to the variables $(x,e)$, and let $L$ be the Lipschitz constant such that 
\be\label{lipscht.def}
\|f(x, k(x+e))\|\le L(\|x\|+\|e\|), \; \textrm{ for all } (x,e)\in \mathcal{X}\times \mathcal{E}. 
\ee 
The first fact that we recall is that as far as $x\in \mathcal{X}$, then $e\in \mathcal{E}$. Therefore, denoted by $\sigma$ the positive constant whose inverse is the Lipschitz constant of the function  $\alpha_1^{-1}(\frac{1}{\lambda(1-c)}\gamma_2(4\|e\|))$ on $\mathcal{E}$ (under the assumption that the function is locally Lipschitz), we have that 
\be\label{lipschitz.err.def}
\alpha_1^{-1}(\frac{1}{\lambda(1-c)}\gamma_2(4\|e\|)) \le \frac{1}{\sigma} \|e\|,\quad e\in \mathcal{E}.
\ee  
Hence, as far as $x(t)\in \mathcal{X}$, any sequence $\{t_k\}$ generated by the triggering rule 
\be\label{cur.mod}
\| e(t) \| \le \sigma \| x(t)\|, \quad t\in [t_k, t_{k+1}]
\ee
implies that (\ref{cur}) is satisfied as well, that is 
\[
\gamma_2(4\| e(t) \|) \le \lambda(1-c) \| x(t)\|, \quad t\in [t_k, t_{k+1}].
\]
Now we prove that with a triggering rule as in (\ref{cur.mod}), the requirement $t_{k+1}-t_k\ge \varepsilon$ for all $k\in \mathbb{N}$ is actually met. To show this, we study the evolution of $e(t)$ for $t\in [t_k, t_{k+1}[$. Observe that 
\[
\dot e(t) = - f(x, k(x+e))
\]
implies 
\[
e(t)\le -\displaystyle\int_{t_k}^t f(x(s), k(x(s)+e(s)))ds
\]
and hence
\[
\|e(t)\|\le \displaystyle\int_{t_k}^t L(\|x(s)\|+\|e(s)\|)ds.
\]
Since $\|x(s)\|\le \|x(t_k)\| + \|e(s)\|$, we also have
\[
\|e(t)\|\le L \|x(t_k)\| (t-t_k) +\displaystyle\int_{t_k}^t 2L \|e(s)\|ds.
\]
Applying Gronwall-Bellman's inequality and after standard but lengthy manipulations, one arrives at 
\[
\|e(t)\|\le \displaystyle\frac{1}{2} \left(
{\rm e}^{2L(t-t_k)} -1
\right)\|x(t_k)\|.
\]
Let $g(\tau):= \frac{1}{2} \left(
{\rm e}^{2L\tau} -1
\right)$.  
Since $\|x(t_k)\| \le \|e(t)\|+\|x(t)\|$, for all $t\ge t_k$ such that  $g(t-t_k)<1$, then the previous inequality is equivalent to 
\[
\|e(t)\|\le \displaystyle\frac{g(t-t_k)}{1-g(t-t_k)}  \|x(t)\|.
\]
Now, 
\[
g(t-t_k)<1\textrm{ and }\displaystyle\frac{g(t-t_k)}{1-g(t-t_k)}  \le \sigma
\]
if and only if 
\[
{\rm e}^{2L(t-t_k)} < \min\{3, \displaystyle\frac{3\sigma+1}{\sigma+1}\}
\] 
or equivalently 
\[
t-t_k < \displaystyle\frac{1}{2L}\ln \left(\displaystyle\frac{3\sigma+1}{\sigma+1}\right).
\]
Thus condition  (\ref{cur.mod}) is not violated until at least 
\[
\displaystyle\frac{1}{2L}\ln \left(\displaystyle\frac{3\sigma+1}{\sigma+1}\right)
\] 
units of time have elapsed. In other words, any sequence of sampling times generated by the triggering rule (\ref{cur.mod}) or (\ref{cur}) occurs at a finite sampling rate, that is  (\ref{minimal_IET}) holds, with $\varepsilon$ replaced by $\varepsilon_R$ and  $\varepsilon_R = \frac{1}{2L}\ln \left(\displaystyle\frac{3\sigma+1}{\sigma+1}\right)$. 

The discussion allows us to draw the following conclusion: 

\begin{theorem}
Consider system (\ref{system}) in closed loop with the control (\ref{actual_control}), under DoS attacks satisfying Assumption  \ref{ass:DoS_slow}, and with control update rule defined in (\ref{cur}). 
Let  Assumption \ref{assumpt2} hold. 
Then, given any $R>0$, if the parameter $\tau$ in (\ref{DoS_slow}) satisfies
\begin{equation} \label{tau_gb2_mod}
\tau \, > \,  \frac{c\lambda}{\lambda+2\mu}\left( 1+ \frac{\Delta_*}{\tau_*} \, \right),  
\end{equation}
where
\begin{equation} \label{Delta_*}
\Delta_* \, := \, \sup_{n \in \mathbb N}  \Delta_{\mathbb S_n}
\end{equation}
and 
\begin{equation} \label{tau_*}
\tau_* \, := \, \inf_{n \in \mathbb N}  \tau_{n}>0,
\end{equation}
then  the inequality 
\be\label{nonl.bound.x}
\| x(t)\|\le \alpha_1^{-1}\left(\gamma {\rm e}^{-\beta t} \alpha_2(\| x(0)\|)
\right)
\ee
holds for all $t\ge 0$ and for all $\|x(0)\|\le R$, where
\[
\gamma =e^{\kappa (\lambda+2\mu)\left( 1+ \frac{\Delta_*}{\tau_*} \, \right)},\quad  \beta= \left(c\lambda- \frac{\lambda+2\mu}{\tau}\left(1+ \frac{\Delta_*}{\tau_*}\right)\right),
\]
and the parameters $\kappa, \lambda, \mu$,  are as in (\ref{DoS_slow}), (\ref{iss}), (\ref{bound.assumpt2}), respectively. 
Moreover, the control update rule has a semi-global uniform finite sampling rate and (\ref{minimal_IET}) holds for each $k\in \mathbb{N}$ with with $\varepsilon$ replaced by $\varepsilon_R$ and  $\varepsilon_R = \frac{1}{2L}\ln \left(\frac{3\sigma+1}{\sigma+1}\right)$. \hfill $\Box$
%
%
\end{theorem}

There is a clear trade-off between the sampling rate and the DoS attacks that the system can tolerate. In fact, the minimum inter-sampling time $\varepsilon_R$ depends on both $\sigma$ and $L$.  In turn, these parameters depend on the radius of the compact set $\mathcal{X}$. This radius is affected by the parameter $\gamma$: the larger is $\gamma$ the larger is the radius. Now, $\gamma$ grows with $\kappa$ that appears in the description of the DoS signals, modeling possible occurrence of large intervals of DoS attacks at initial times. Thus, a sustained attack at initial times might lead to a small minimum  inter sampling-time.  On the other hand the parameter $\tau$ that accounts for the percentage of time for which an attack takes place depend on both the data of the nonlinear system under control and the characteristic of the DoS attack signal. \vspace{0.1cm}

The result gives an indication on how to implement event-triggered control to stabilize nonlinear systems in the presence of DoS attacks. Event-triggered control requires continuous monitoring of the state $x(t)$ and can be resource-consuming. Alternative implementations in the spirit of Section 4.2 in \cite{CDP:PT:ARXIV} will be investigated in future versions of this work.

\section{Conclusions} \label{sec:conclusions}

The paper investigates the design of event-based control strategies for nonlinear systems in the presence of DoS attacks that interrupts the flow of information from the sensors to the actuators. The DoS signal attack is modeled at a fairly general level that we believe allows for the inclusion of several interesting scenarios. Relations between the sampling frequency, the data of the nonlinear systems under control and the features of the DoS attack signal have been revealed. \vspace{0.1cm}

 The main working assumption is Assumption \ref{assumpt2} whose  role is to prevent the occurrence of finite escape times. It therefore restricts the class of nonlinear systems but allows for a less complicated analysis. Clearly, removing this assumption requires to restrict the class of DoS attacks the system can tolerate: if the systems undergoes prolonged attacks, it will evolve in open loop for long time intervals facing the possible occurrence of a finite escape time. This alternative formulation may be worth of investigation. \vspace{0.1cm}
 
In future work, more attention will be given to the actual implementation of our resilient control and in particular to its connections with other event-based  approaches such as self-triggered control. Relevant case studies to assess the effectiveness of our approach are also part of our future research plan. Whether our method can deal with  attack   scenarios different from DoS attacks is a topic worth of investigation as well. \vspace{0.1cm}

 Robustness of the proposed resilient control (as well as of its linear counterpart studied in \cite{CDP:PT:IFAC14,CDP:PT:ARXIV}) to external disturbances in an ISS sense is a very interesting and challenging research topic that will be tackled in the future. \vspace{0.1cm}
 
 Our initial interest for resilient control was motivated by distributed control strategies for dynamical networks in a cyberphysical environment (see e.g.~\cite{CDP:PF:TAC13}). We believe that our technique can be extended to distributed resilient control and can be a very fertile research ground.

\section{Appendix}

\textbf{Proof of Lemma \ref{lem:tech:nonl}}
Recall that $e(t) \, = \, x(t_{k(h_n)}) - x(t)$ by definition, so that
\[\ba{c}
\| x(t_{k(h_n)})\| -  \| x(h_n)\| \le \| e(h_n) \|\\
\le \displaystyle \frac{1}{4} \gamma_2^{-1}\left(\lambda(1-c) V(x(h_n))\right)
\ea\]
where the second inequality descends from the control update rule (\ref{cur}). Hence
\[\ba{c}
\| x(t_{k(h_n)})\|    \le \displaystyle \frac{1}{4} \gamma_2^{-1}\left(\lambda(1-c) V(x(h_n))\right) + \| x(h_n)\|. 
\ea\]
On the other hand
\[
\| x(h_n)\| \le \alpha_1^{-1} (V(x(h_n)))\le \displaystyle\frac{1}{4}   \gamma_2^{-1}(\mu V(x(h_n))),
\] 
where the second inequality is implied by condition (\ref{bound.assumpt2}). Combining  the two inequalities above the thesis descend. 
\hfill \qedp

\vspace{0.2cm}

\textbf{Proof of Theorem \ref{thm:gb2}}
For any $t\in \mathbb{R}_{\ge 0}$, the Lyapunov function evolves as $\dot V(x(s))\le -\omega_1 V(x(s))$ if $s\not \in \Xi(t)$   and 
$\dot V(x(s))\le \omega_2 V(x(s))$ if $s\in \Xi(t)$. Hence
\[\ba{rcl}
V(x(t)) &\le & {\rm e}^{-\omega_1(t-\left| \Xi(t)\right|)} {\rm e}^{\omega_2 \left| \Xi(t)\right|}  V(x(t_0))\\
&\le & {\rm e}^{-\omega_1t + (\omega_1+\omega_2)\left| \Xi(t)\right|}  V(x(t_0)).
\ea\]
Bearing in mind Assumption \ref{ass:DoS_slow}, and in particular condition (\ref{DoS_slow}), it follows that 
\[\ba{rcl}
V(x(t)) &\le &  {\rm e}^{-(\omega_1- (\omega_1+\omega_2)\tau) t + \kappa (\omega_1+\omega_2)}  V(x(t_0)).
\ea\]
The thesis now follows immediately under the stated assumptions. 
\hfill \qedp

\bibliographystyle{plain} 
\bibliography{Automatica_PAPER_BIBLIOa}

\end{document}